\documentclass[twocolumn,showpacs,amsmath,amssymb,aps,prl,nofootinbib]{revtex4-1}

\newcommand{\tl}{\tilde \lambda}
\newcommand{\bk}{{\bf k}}
\newcommand{\bq}{{\bf q}}

\newcommand{\beq}{\begin{eqnarray}}
\newcommand{\eeq}{\end{eqnarray}}
\newcommand{\beqq}{\begin{eqnarray*}}
\newcommand{\eeqq}{\end{eqnarray*}}

\newcommand{\be}{\begin{equation}}
\newcommand{\ee}{\end{equation}}

\newcommand{\tL}{\tilde \Lambda}

\newcommand{\bp}{{\bf p}}
\newcommand{\vf}{v_F}

\newcommand{\ve}{{\varepsilon}}

\usepackage{textcomp}
\usepackage[colorlinks=true,citecolor=blue,linkcolor=blue]{hyperref}
\usepackage{graphicx}
\usepackage{dcolumn}
\usepackage{bm}
\usepackage{color}
\usepackage{tabularx}

\usepackage{filecontents}

\usepackage{amsmath}

\usepackage{dsfont}
\usepackage{changepage}

\usepackage{fancyhdr}
\usepackage{amsthm, amssymb}
\usepackage{floatflt}
\usepackage{float}
\usepackage{array}
\usepackage{blindtext}

\begin{document}

\begin{titlepage}

\title{Thermal plasmon resonantly enhances electron scattering in Dirac/Weyl semimetals}

\author{Vladyslav Kozii}
\thanks{Correspondence should be addressed to koziiva@mit.edu}
\affiliation{Department of Physics, Massachusetts Institute of Technology, Cambridge,
Massachusetts 02139, USA}

\author{Liang Fu}

\affiliation{Department of Physics, Massachusetts Institute of Technology, Cambridge,
Massachusetts 02139, USA}

\date{\today}

\begin{abstract}
We study the inelastic scattering rate due to the Coulomb interaction in three-dimensional Dirac/Weyl semimetals at finite temperature. We show that the perturbation theory diverges because of the long-range nature of the interaction, hence, thermally induced screening must be taken into account. We demonstrate that the scattering rate has a non-monotonic energy dependence with a sharp peak owing to the resonant decay into thermal plasmons. We also consider the Hubbard interaction for comparison. We show that, in contrast to the Coulomb case, it can be well described by the second-order perturbation theory in a wide energy range.

\end{abstract}

\pacs{}

\maketitle

\draft

\vspace{2mm}

\end{titlepage}

Three-dimensional Dirac semimetals have attracted great interest in the condensed matter community due to their exotic electronic properties~\cite{AbrikosovBeneslavskii1970,Abrikosov98,Murakami07,KoshinoAndo10,Wanetal11,Yangetal11,Xuetal11,BurkovBalents11,Liuetal14,Potteretal14,Neupaneetal15,Xuetal15,Lvetal15,HosurParameswaranVishwanath2012,RosensteinLewkowicz2013,Armitage18,Burkov18}. The low-energy excitations of these materials are massless Dirac fermions with a linear dispersion near the touching points between the conductance and valence bands. If the Kramers degeneracy of the Dirac cones is removed by breaking either time reversal or inversion symmetry, a topological Weyl semimetal (WSM) is realized~\cite{Wanetal11,Murakami07}. Weyl nodes are monopoles of Berry curvature in momentum space, hence, they are topological objects and can be eliminated only by merging with another node of opposite monopole strength.



Although non-interacting WSMs are already intriguing due to their nontrivial topological properties, the interaction effects in these materials are of great interest. In particular, inelastic electron-electron scattering is expected to be crucial for determining the conductivity \cite{HosurParameswaranVishwanath2012,RosensteinLewkowicz2013} and spectral properties~\cite{HofmannBarnesDasSarma2015,ThrockmortonHofmannBarnesDasSarma2015,SetiawanDasSarma2015} of clean samples at low temperatures, which can be directly probed in transport and angle-resolved photoemission spectroscopy (ARPES)/scanning tunneling microscopy (STM) measurements, respectively.



While most of the previous studies of the spectral function were focused on the zero-temperature case, certain interesting phenomena are expected in interacting WSMs at finite temperature. For example, finite-lifetime quasiparticles can display novel spectral features described by the non-Hermitian topological theory~\cite{Esakietal2011,Liangetal2013,Lee2016,Leykametal2017,MenkeHirschmann2017,Xuetal2017,ShenZhenFu2017,KoziiFu2017,GonzalesMolina2017,Huetal2017,ZyuzinZyuzin2017,Michal2018,Huitao2018}. A call for a profound  understanding  of these intriguing  phenomena that can be measured in ARPES experiments   motivates us to study the electron's self-energy in WSMs at finite temperature.

In this paper we focus on the inelastic quantum scattering rate (inverse quasiparticle's lifetime)
due to the electron-electron interaction. We consider the cases of the Coulomb and  repulsive Hubbard (short-range) interactions. We find that the second-order perturbation theory generically diverges in the case of a Coulomb interaction, and a summation of an infinite series of diagrams within the random phase approximation (RPA) is required. At finite temperature, the collective density oscillations of thermally excited carriers can be considered as thermal plasmons. We show that the thermally induced screening and thermal plasmons lead to a strong energy dependence of the electron scattering rate which exhibits a sharp peak around the plasma frequency $\omega_{\text{pl}} \propto T$. This peak can be viewed as a consequence of a strong electron-plasmon interaction. Although we do not aim to describe any specific experiment in our study, we believe it shares similar physics with certain features that were attributed to the coupling between electrons and plasmons and were observed in optical measurements in elemental bismuth \cite{Tediosi2007} and $\text{Na}_3\text{Bi}$ \cite{Jenkins2016}. Additionally, while we focused on the case of Dirac semimetals in our work, we believe that the same physics is relevant for other semimetallic systems with a low carrier density. For example, we consider half-Heusler compounds with quadratic band touching as promising candidates  for testing our findings \cite{Nakajima2015}.

Among other results, we find that the scattering rate vanishes logarithmically at exponentially small energies. We also show that the model with the Hubbard interaction, in contrast to the Coulomb case, allows for a perturbative calculation of the scattering rate in a wide range of energies. At the smallest energies, however, it also approaches zero in a nonanalytic way. We hope that our results can be directly probed by measuring the spectral function in ARPES experiments.





{\it Model.} We consider a model for WSM at a neutrality point with $N$ identical isotropic Weyl nodes. The low-energy Hamiltonian in the presence of an interaction has the form $H = H_0 + H_{\text{int}}$, with

\begin{align}
&H_0 = \sum_{i, \bk}  \chi_i \vf \psi^\dagger_{\bk, i s} \bk \cdot \boldsymbol{\sigma}_{ss'} \psi_{\bk, i s'}, \nonumber \\ &H_{\text{int}} = \frac12 \sum_{\bk, \bp, \bq} \psi^\dagger_{\bk - \bq, i s} \psi_{\bk, i s} V_0(\bq) \psi^\dagger_{\bp + \bq, j s'} \psi_{\bp, j s'}. \label{Eq:H}
\end{align}
Here, $\psi_{\bk, i s}$ is a two-component spinor in the pseudospin space $s$, $\boldsymbol{\sigma}$ is a vector of Pauli matrices, $i, j = 1,\ldots, N$ numerate Weyl nodes, $\chi_i = \pm 1$ is the chirality of the $i$th node, and $\vf$ is the Fermi velocity. A summation over repeating indices is implied. The bare Coulomb interaction is given by $V_0(\bq) = 4\pi e^2/\epsilon q^2$, where $\epsilon$ is a dielectric constant of a material, and the repulsive Hubbard interaction is described by $V_0(\bq) = \lambda >0$. In what follows, we neglect the internodal scattering as well as the nonzero curvature of a single-electron spectrum, which, in principle, can play an important role at small energies~\cite{PaakseKhveshchenko2000,ChubukovTsvelik}, leaving these questions for future study. We use units with $\hbar = k_B = 1$ throughout the paper.

{\it The Coulomb interaction.} The strength of the Coulomb interaction in Weyl materials is measured by the dimensionless effective fine-structure constant, $\alpha = e^2 / \epsilon \vf$, which is a density-independent ratio of a typical Coulomb energy to kinetic energy. In this paper, we only consider the case of a weak interaction, which is a reasonable assumption for some real materials with a large dielectric constant. For example, fine-structure constants for $\text{Bi}$, $\text{Na}_3\text{Bi},$ and $\text{Cd}_3\text{As}_2$ can be estimated to be $\alpha_{\text{Bi}} \le 0.2$, $\alpha_{\text{Na}_3\text{Bi}}\approx 0.15,$ and $\alpha_{\text{Cd}_3\text{As}_2}\approx 0.04$ \cite{Tediosi2007,Zhuetal2011,Ruhman2017,Jenkins2016,Liangetal2015,Liuetal14,Jenkins2016,Young73,Skinner2014}. To analytically control our calculation, we further require a large number of Weyl nodes, $N\gg1$, but keep the product $\alpha N \ll 1$ small. Finally, we also assume that the more restrictive condition is satisfied, $\alpha N \ln (\vf \Lambda/ T) \ll 1,$  where $\Lambda$ is a high-momentum cutoff of the order of the distance between nodes. The latter assumption can be easily relaxed and is used here only to simplify some formulas. Despite the approximations made above, we expect our results to be qualitatively correct even for an interaction strength of order one.


Before we consider the inelastic scattering rate, we briefly comment on the velocity and fine-structure renormalization due to the Coulomb interaction. This question was studied, e.g., in Refs.~\onlinecite{AbrikosovBeneslavskii1970,HosurParameswaranVishwanath2012,Isobe2012}. It was found that the Fermi velocity and fine-structure constant at the scale of temperature $T$ are renormalized to the leading order as $\vf(T) = \vf(\alpha_0/\alpha_T)^{2/N+2}$ and $\alpha_T = \alpha_0\left[1+\frac{(N+2)\alpha_0}{3\pi}\ln(\vf\Lambda/T)\right]^{-1},$ where $\vf$ and $\alpha_0$ are bare values at the scale $\vf \Lambda$. We use the renormalized parameters hereafter.

The nonzero scattering rate results from the imaginary part of the interaction potential. Since the bare Coulomb interaction is real, we need to take into account the screening effects, e.g., within the RPA. The effective interaction then has the form

\be
V^R(\omega, \bq) = \frac{V_0(\bq)}{1+V_0(\bq)N \Pi^R(\omega, \bq)}, \label{Eq:V}
\ee
where $\Pi^R(\omega, \bq)$ is a polarization operator. Generally, the RPA is justified in the limit of a large number of Weyl nodes, $N\gg1$; however, as discussed in Ref.~\onlinecite{Mirlin2011}, at finite temperature the RPA is valid even at $N\sim1$ due to the thermally induced screening, provided relevant momenta satisfy the condition $\vf q \lesssim T.$

While the evaluation of the polarization operator at $T=0$ is straightforward~\cite{LvZhang2013,Zhou2015}, the calculation  at finite temperature is a very complicated task that can usually be accomplished only numerically. Nevertheless, following the method used in Ref.~\onlinecite{Mirlin2011}, we find an approximate analytical expression for $\Pi^R$ in the most relevant limiting cases~\cite{SM},

\begin{widetext}
\be
\Pi^R( \Omega, Q) = \frac{T^2}{\vf^3}\left\{  \begin{array}{lc} \frac16\left( 1 - \frac{|\Omega|}{2Q}\ln\frac{|\Omega|+Q}{|\Omega|-Q}  \right) + \frac{Q^2}{3\pi^2}\ln \frac{\tilde \Lambda}{\max \{ 1, |\Omega| \}} +i \frac{Q^2}{6\pi}\tanh\frac{\Omega}2, & Q\ll1, \, Q<|\Omega|\\ \frac16\left( 1 - \frac{|\Omega|}{2Q}\ln\frac{Q + |\Omega|}{Q - |\Omega|}  \right) + \frac{Q^2}{3\pi^2}\ln \tilde \Lambda + i\frac{\pi}{12}\frac{\Omega}Q, & Q\ll1, \, Q>|\Omega| \\ \frac{Q^2}{3\pi^2}\ln \frac{\tilde \Lambda}{\sqrt{\Omega^2 - Q^2}} + i \frac{Q^2}{6\pi}\text{sign} \Omega, & Q\gg1, \, Q<|\Omega| \\ \frac{Q^2}{3\pi^2}\ln \frac{\tilde \Lambda}{\sqrt{Q^2 - \Omega^2 }} + i \frac1{\pi} e^{-Q} \sinh\Omega , & Q\gg1, \, Q>|\Omega| \end{array}     \right. \label{Eq:Pi}
\ee
\end{widetext}
where we defined the dimensionless quantities $Q \equiv \vf q /2T$, $\Omega \equiv \omega/2T,$ and $\tilde \Lambda \equiv \vf \Lambda /2T$. In the zero-temperature limit, $Q \gg 1$, we reproduce the result by Abrikosov and Beneslavski\u{\i}~\cite{AbrikosovBeneslavskii1970}, $\Pi(\omega, q) = \frac{q^2}{12 \pi^2 \vf}\ln \frac{\Lambda}{\sqrt{q^2-\omega^2/\vf^2}}.$

In the static limit, $\omega = 0$, the polarization operator~(\ref{Eq:Pi}) determines the thermally induced screening of the Coulomb potential, and the effective interaction at low momenta takes the form

\be
V(\omega = 0, \vf q \ll T) = \frac{4 \pi \alpha \vf}{q^2 + l^{-2}_{\text{scr}}},
\ee
where the screening length is given by $l_{\text{scr}}^{-1} = \frac{T}{\vf}\sqrt{\frac{2\pi}3 \alpha N}.$

\begin{figure*}
\includegraphics[width=1.\textwidth]{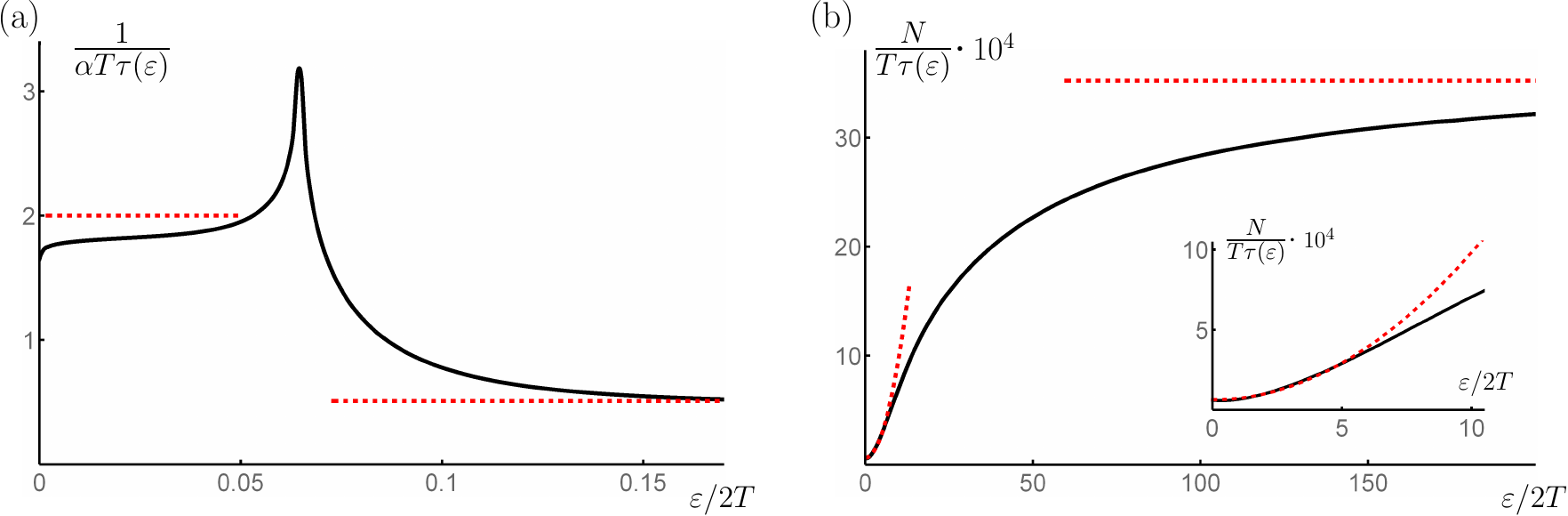}
\caption{The inelastic scattering rate as a function of energy due to (a) Coulomb and (b) Hubbard interactions. Red dashed lines correspond to the asymptotic analytical expressions given by Eqs.~(\ref{Eq:taucoulomb}) and (\ref{Eq:tauhubbard}). At exponentially small energies, the scattering rate logarithmically approaches zero in both cases  (not displayed in this figure). (a) The scattering rate exhibits nonmonotonic behavior with a sharp peak at $\ve = \omega_{\text{pl}}/2$ owing to thermal plasmons. The coupling constant equals $\alpha N = 0.1$. (b) Coupling constant $\lambda$ is such that $\lambda N T^2/\vf^3 = 0.03.$ The inset shows that the scattering rate is well described within the second-order perturbation theory (dashed line) in a wide range of energies around $\ve \approx T$. At higher energies, however, the perturbative result smoothly crosses over to a constant, which can only be obtained after the RPA summation. This crossover occurs only for extremely large values of the ultraviolet cutoff $\Lambda$ satisfying $\Lambda \gg \sqrt{\vf/\lambda N \ln(\vf \Lambda/T)}$.}
\label{Fig}
\end{figure*}

In the region $\vf q \ll \omega \ll T$, the real part of the polarization operator becomes negative, giving rise to thermally induced plasmon excitations~\cite{Hofmann2015,Kharzeev2015}. At low momenta, the plasmon dispersion is determined by the equation $1 + N V_0(q) \Pi(\omega \gg \vf q) = 0,$ yielding the solution

\begin{align}
&\omega  = \omega_{\text{pl}} + \frac{3}{10} \frac{\vf^2 q^2}{\omega_{\text{pl}}} - i \Gamma, \nonumber \\
&\omega_{\text{pl}} = T\sqrt{\frac{2\pi}9 \alpha N} \ll T, \quad \Gamma = \frac{3}{32\pi} \frac{\omega_{\text{pl}}^4}{T^3} \ll \omega_{\text{pl}}.
\end{align}
At the neutrality point, the only energy scale is set by temperature, hence, it is natural that $\omega_{\text{pl}}\propto T$. We stress that,  at weak coupling, the damping of thermal plasmons in WSMs is small compared to their energy, consequently, they are well-defined collective excitations.

To study the inelastic scattering rate, we calculate the imaginary part of the electron's self-energy, $\text{Im} \, \Sigma (\omega, \bk),$ at finite temperature. As discussed in Ref.~\onlinecite{Mirlin2011} in the context of graphene, the electron's self-energy is generally a matrix in the pseudospin basis and can be parametrized as $\Sigma(\ve,\bk) = \Sigma_{\ve} I + \Sigma_v \boldsymbol{\sigma} \cdot \hat \bk.$ It is natural to associate the scattering rate with $\text{Im}\, \Sigma_{\ve}$ taken on the mass shell, in the spirit of the conventional Fermi liquid (FL),

\be
 \frac1{2\tau(\ve)} \equiv  \left. - \text{Im} \, \Sigma^R_{\ve} (\ve, \bp)\right\vert_{p = |\ve|/\vf}.
\ee
It is clear that $\tau(\ve) = \tau(-\ve)$  at the neutrality point due to particle-hole symmetry, so we focus on positive energies hereafter.



In the one-loop approximation, the imaginary part of the electron's self-energy reads as~\cite{AGD}

\begin{multline}
\text{Im} \, \Sigma^R_{\ve}(\ve,\bk) = \frac14 \sum_{i = \pm} \int \frac{d^3q}{(2\pi)^3} \text{Im} V^R(\omega_i, \bq)\\ \times \left(\coth\frac{\omega_i}{2T} + \tanh\frac{\ve - \omega_i}{2T}  \right),
\end{multline}
where we defined $\omega_\pm \equiv \ve \pm \vf |\bk - \bq|.$ After a straightforward but rather cumbersome calculation, we find~\cite{SM}

\be
\frac1{\tau(\ve)} = \left\{\begin{array}{ll} c_1 \frac{T }N  \left(\ln\frac{T}{\ve}\right)^{-2}, \qquad & \ve \ll T\exp\left( - \frac{c_2}{ N \alpha}  \right),  \\  2\alpha T, \qquad &  T\exp\left( - \frac{c_2}{ N \alpha}  \right) \ll \ve \ll T\sqrt{\alpha N} ,\\ \frac34 \alpha T \ln \frac1{\alpha N}, \qquad &  \ve = T\sqrt{\frac{\pi N \alpha}{18}} = \frac{\omega_{\text{pl}}}2,
 \\  0.55 \alpha T, \qquad & \ve \gg T\sqrt{\alpha N}  ,   \end{array}  \right. \label{Eq:taucoulomb}
\ee
where $c_1$ and $c_2$ are numerical coefficients of order 1. The main contribution to the first region comes from the bosonic frequencies and momenta of order of temperature, which are not accurately captured by Eq.~(\ref{Eq:Pi}), hence, coefficient $c_1$ cannot be calculated within our approach.


The behavior of the scattering rate as a function of energy is shown in Fig.~\ref{Fig}(a). We emphasize that even though we consider a weak-coupling limit, result~(\ref{Eq:taucoulomb}) is nonperturbative. Indeed, the naive lowest-order weak-coupling answer (a single polarization bubble in the effective interaction) would be proportional to $\tau^{-1}_{\text{naive}} \propto \alpha^2 N ,$ which does not hold in any of the energy domains. This is due to the singular form of the Coulomb interaction at low momenta, which eventually leads to the infrared divergence and requires the RPA resummation.  This conclusion is similar to the result for a two-dimensional (2D) analog of the problem, graphene, considered in Ref.~\cite{Mirlin2011}.

\begin{table*}[t]
\begin{center}
\scalebox{1.02}{
\begin{tabularx}{0.983\textwidth}{| c | c | c | c | c | c|}
    \hline {\bf Coulomb}  & $\ve, \xi_{\bk} \ll T e^{-c_2/\alpha N}$ & $ Te^{-c_2/\alpha N}\ll\ve, \xi_{\bk} \ll T\sqrt{\alpha N}$ & $\ve, \xi_{\bk} = \omega_{\text{pl}}$ & $T\sqrt{\alpha N} \ll \ve, \xi_{\bk} \ll T$ &  $\ve, \xi_{\bk} \gg T$ \\ \hline
   $\text{Im}\,\Sigma_{\ve} (\ve,0)$ & \scalebox{1.35}{$\sim \frac{T}{N\ln^2 (T/\ve)}$} & $\alpha T$ & $\alpha T/2$& \scalebox{1.35}{$\frac{2\pi \alpha^2 N T^3}{3\ve^2}$} & \scalebox{1.35}{$\frac{\alpha^2 N \ve }{12\pi}$} \\ \hline
  $\text{Im}\,\Sigma_{\ve}(0, \bk)$ &\scalebox{1.35}{$\sim  \frac{T}{N\ln^2 (T/\xi_{\bk})}$} & $\alpha T$ & \scalebox{1.35}{$\frac34$}$\alpha T \ln (1/\alpha N)$ & \scalebox{1.35} {$\frac{2\pi \alpha^2 N T^3}{9\xi_{\bk}^2}$ }& \scalebox{1.35} {$\frac{2\alpha^2 N T^2 \exp(-\xi_{\bk}/2T)}{3\pi \xi_{\bk}}$} \\
    \hline
\end{tabularx}
}
\end{center}

\begin{center}
\scalebox{1.03359}{
\begin{tabularx}{.971\textwidth}{| c | c | c | c | c |}
    \hline {\bf Hubbard} & $\ve, \xi_{\bk} \ll T\exp\left( - \frac{b_2 \vf^3}{ \lambda N T^2}\right)$& $ T\exp\left( - \frac{b_2 \vf^3}{ \lambda N T^2}  \right)\ll\ve, \xi_{\bk} \ll T$ & $T \ll \ve, \xi_{\bk} \ll \sqrt{\frac{\vf^3}{\lambda N \ln (\vf \Lambda / T)}}$ & $ \ve, \xi_{\bk} \gg \sqrt{\frac{\vf^3}{\lambda N \ln (\vf \Lambda / T)}} $ \\ \hline
{$\text{Im}\,\Sigma_{\ve} (\ve,0)$}&\scalebox{1.35}{$\sim \frac{T}{N\ln^2 (T/\ve)}$}& 0.035\scalebox{1.35}{$\frac{\lambda^2 N T^5}{\vf^6}$} & \scalebox{1.35}{ $ \frac{\lambda^2 N \ve^5}{15360 \pi^3 \vf^6}$ }& \scalebox{1.35}{ $\frac{3\pi\ve}{4N\ln^2(\vf \Lambda/\ve)}$ }\\ \hline
 {$\text{Im}\,\Sigma_{\ve}(0, \bk)$}& \scalebox{1.35}{$\sim  \frac{T}{N\ln^2 (T/\xi_{\bk})}$} & 0.035\scalebox{1.35}{$\frac{\lambda^2 N T^5}{\vf^6}$} & \scalebox{1.35}{ $\frac{\lambda^2 N T^2 \xi_{\bk}^3 \exp(-\xi_{\bk}/2T)}{384\pi^3\vf^6}$ } & \scalebox{1.35}{$\frac{6\pi T^2 \exp(-\xi_{\bk}/2T)}{N\xi_{\bk} \ln^2(\vf \Lambda/\sqrt{\xi_{\bk} T})}$} \\
    \hline
\end{tabularx}
}
\end{center}
\caption{ The imaginary part of the electron's self-energy in the limits of zero energy or momentum due to the Coulomb (top) or the Hubbard (bottom) interactions. The single-electron spectrum is defined as $\xi_{\bk} \equiv \vf k$. In the case of the Coulomb interaction, $\text{Im} \, \Sigma_{\ve}(0, \bk)$ exhibits a strong peak at $\xi_{\bk} = \omega_{\text{pl}}$ due to resonant excitation of the thermal plasmons. }
\label{Table:ImSigma}

\end{table*}

At small energies, $\ve \lesssim 2\alpha T$, the quasiparticles are not well defined, since $\tau^{-1}(\ve) > \ve$ in this range. The scattering rate of these states, however, is determined by electrons with higher energies, which ensures the self-consistency of our calculation. This is in analogy with the conventional FL theory at nonzero temperature, where the finite lifetime of quasiparticles at vanishing energy is determined by thermal excitations. There is, however, an interesting difference between WSM and FL at exponentially small energies, $\ve \ll T \exp(-c_2/\alpha N).$ In this regime, the scattering rate in FL saturates to a constant value, $\tau_{\text{FL}}(\ve \to 0) \propto T^2$, while in WSM it logarithmically approaches zero, see Eq.~(\ref{Eq:taucoulomb}). The reason for such behavior is rooted in the logarithmical divergence of the real part of the polarization operator~(\ref{Eq:Pi}) at $|\omega| \approx \vf q \sim T$. For exponentially small energies $\ve$, one has $\ln||\Omega| - Q| \sim \ln (\ve/T),$ which eventually determines $\tau^{-1} \propto \ln^{-2}(T/\ve)$ dependence in this regime.

As the energy of the quasiparticles increases, the scattering rate exhibits a non-monotonic behavior. In particular, it has a sharp logarithmically enhanced peak at $\ve = \omega_{\text{pl}}/2$ due to the resonant excitation of the thermal plasmons~\cite{SM}. This distinctive feature is exclusive for 3D and absent in graphene~\cite{Mirlin2011}. At zero temperature, the scattering rate is zero because of phase-space restrictions~\cite{HofmannBarnesDasSarma2015}. Since $\tau^{-1}\sim \alpha T$ for most of energies, WSM with the Coulomb interaction can be called a marginal FL.

Next, we calculate the self-energy in two other important limits, $\text{Im}\, \Sigma_{\ve} (\ve,\bk=0)$ and $\text{Im}\, \Sigma_{\ve}(\ve = 0, \bk)$, and present our results in Table~\ref{Table:ImSigma}. We see that in the region $\ve \ll T\sqrt{\alpha N}$ the answer is non-perturbative and coincides (up to a possible numerical prefactor) with the scattering rate, Eq.~(\ref{Eq:taucoulomb}). At higher energies, on the contrary, one can use second-order perturbation theory. We also notice that the plasmon peak is absent in $\text{Im}\, \Sigma_{\ve} (\ve,\bk=0),$ because the plasmon resonance cannot be achieved in this case due to frequency-momentum mismatch~\cite{SM}.

Formally, the second-order perturbative result for $\text{Im} \, \Sigma_{\ve} (\ve,\bk=0)$ and $\text{Im}\, \Sigma_{\ve}(\ve = 0, \bk)$ converges (no infrared divergence in momentum integral) for any non-zero $\ve \ne 0$ or $\xi_{\bk} = \vf k \ne 0$. However, upon decreasing the energy of the excitations, it grows as $1/\ve^2$ due to processes with small momenta transfer, signalizing that the naive perturbation theory becomes insufficient and a full summation of the most divergent terms is required~\cite{PaakseKhveshchenko2000,ChubukovTsvelik}. After summation within the RPA, the $1/\ve^2$ behavior crosses over to a physically meaningful non-perturbative result at small energies, $\ve,\xi_{\bk} \lesssim T\sqrt{\alpha N}$, as shown in Table~\ref{Table:ImSigma}.

{\it The Hubbard interaction.} Now we perform a similar analysis for the case of a repulsive Hubbard interaction, which may be relevant for certain cold-atom systems. Since we neglect internodal scattering for simplicity, our model is described by the same Hamiltonian~(\ref{Eq:H}) with $V_0(\bq) = \lambda$. Again, we assume a weak-coupling limit and large-$N$ approximation to justify the RPA summation where needed. Specifically, we focus on small coupling constants $\lambda$ satisfying $\lambda N T^2 \vf^{-3} \ll 1$. Furthermore, analogously to the case of the Coulomb interaction, we impose a more restrictive condition, $\lambda N T^2 \vf^{-3} \ln (\vf \Lambda/ T) \ll 1,$ in order to simplify the final expressions.



Similarly to what we found before, finite temperature generates stable collective excitations. In the case of Hubbard repulsion, those are zero-sound modes, with the dispersion determined by the equation $1 + \lambda N \Pi(\omega, \bq) = 0.$ In the low-frequency limit, $\vf q < \omega\ll T$, we find a solution
\begin{align}
&\omega = (\vf + \delta \vf) q - i \Gamma(q), \nonumber \\ &\delta \vf =  \frac{2}{e^2} \vf \exp\left(-\frac{12 \vf^3}{\lambda N T^2}\right) \ll \vf, \nonumber \\
&\Gamma(q) = \frac1{4\pi e^2}\frac{\vf^4 q^4}{T^3} \exp\left(-\frac{12 \vf^3}{\lambda N T^2}\right) \ll \omega.
\end{align}
Since the damping is exponentially small, the zero sound is a well-defined excitation provided its energy is smaller than temperature.



While careful calculation of the scattering rate requires a summation of the infinite RPA series for an effective interaction, it is instructive to first consider the result obtained within the second-order perturbation theory. We find that it gives the answer expected from simple scaling arguments, $\text{Im} \, \Sigma_{\ve} (\ve, \bk) \sim \lambda^2 N \vf^{-6} T^5$ for $\ve, \vf k\ll T$, independently of the ratio $\ve/\vf k$. This is in sharp  contrast to the 2D version of the problem studied in Refs.~\cite{PaakseKhveshchenko2000,ChubukovTsvelik} or the case of a Coulomb interaction studied above, where perturbation theory completely fails at low energies.

Since the perturbative result does not display any dangerous divergencies, it is tempting to conclude that such an approach is sufficient, and no RPA summation is needed at weak coupling. Though this statement is true in a large energy domain, the correct answer obtained within the RPA is more peculiar,

\be
 \frac1{\tau(\ve)}= \left\{\begin{array}{ll} b_1 \frac{T }N \left( \ln \frac{T}{\ve} \right)^{-2}, \qquad & \ve \ll T\exp\left( - \frac{b_2 \vf^3}{ \lambda N T^2}  \right),  \\  0.07 \frac{\lambda^2 N T^5}{\vf^6}, \qquad &  T\exp\left( - \frac{b_2 \vf^3}{ \lambda N T^2}   \right) \ll \ve \ll T ,\\ \frac{3\zeta(3)+4}{96\pi^3}\lambda^2 N \frac{T^3 \ve^2}{\vf^6}, \qquad &  T \ll \ve \ll \sqrt{\frac{\vf^3}{\lambda N \ln (\vf \Lambda / T)}},
\\  \frac3{2\pi}\lambda \frac{T^3}{\vf^3}\left({\ln \frac{\vf \Lambda}{T}} \right)^{-1} , \qquad & \sqrt{\frac{\vf^3}{\lambda N \ln (\vf \Lambda / T)}} \ll \ve \ll \vf \Lambda  .   \end{array}  \right. \label{Eq:tauhubbard}
\ee
Here, $b_{1,2}$ are coefficients of order 1, $\zeta(x)$ is the Riemann zeta-function, and we assumed that $\Lambda^2 \gg \vf / \lambda N \ln (\vf \Lambda/T)$ [in the opposite limit, the last interval in Eq.~(\ref{Eq:tauhubbard}) is absent]. We see that, as anticipated, the second-order perturbation theory is applicable in a wide energy range, failing only for exponentially small, $\ve \ll T\exp\left( - {b_2 \vf^3}/{ \lambda N T^2}  \right)$,  and parametrically large, $\ve \gg \sqrt{{\vf^3}/{\lambda N \ln (\vf \Lambda / T)}}$, energies. From a technical perspective, the reason for the deviation from the perturbative result in these regimes is clear: even though no singularities appear at the second order, a large logarithmical factor shows up in the third order and proliferates with the order of perturbation. Hence, the RPA summation is necessary, resulting in the first and last lines of Eq.~(\ref{Eq:tauhubbard}). The energy dependence of the scattering rate due to the Hubbard interaction is shown in Fig.~\ref{Fig}(b).


The results for  $\text{Im} \, \Sigma_{\ve} (\ve,\bk=0)$ and $\text{Im}\, \Sigma_{\ve}(\ve = 0, \bk)$ are summarized in Table~\ref{Table:ImSigma}, demonstrating again  the relevance of perturbation theory in a big energy interval. Interestingly, the result for the self-energy at zero momentum formally has a simple scaling behavior, $\text{Im} \, \Sigma_{\ve} (\ve,\bk=0) \sim \lambda^2 N \max\{ \ve^5, T^5  \}.$ In practice, however, the prefactor at $T^5$ is four orders of magnitude larger than that at $\ve^5$, which must be taken into account when applied to real materials. An analogous situation was encountered in the study of the relaxation rate in quantum dots in Ref.~\cite{KoziiSkvortsov2016}.


{\it Conclusions.} We studied the scattering rate due to a weak electron-electron interaction in three-dimensional Dirac/Weyl semimetals at finite temperature. We considered the cases of Coulomb and Hubbard interactions. We found that in the Hubbard case the scattering rate can be found within the second-order perturbation theory in a wide range of energies. On the other hand, the Coulomb interaction necessarily requires the RPA summation because of its long-range nature; this results in the non-monotonic sharply peaked energy dependence of the scattering rate due to thermally induced plasmon resonance. In both cases, the scattering rate non-analytically approaches zero at exponentially small energies.


{\it Acknowledgments.} We are very grateful to Max Metlitski for productive discussions, and to Jonathan Ruhman and Cyprian Lewandowski for reading the manuscript and giving valuable feedback. This work was supported by the DOE Office of Basic Energy Sciences, Division of Materials Sciences and Engineering under Award DE-SC0010526.

\widetext
\begin{center}
\textbf{\large Supplemental Material for "Thermal plasmon resonantly enhances electron scattering in Dirac/Weyl semimetals"} 
\end{center}
\setcounter{equation}{0}
\setcounter{figure}{0}
\setcounter{table}{0}

\makeatletter
\renewcommand{\theequation}{S\arabic{equation}}
\renewcommand{\thefigure}{S\arabic{figure}}
\renewcommand{\thetable}{S\Roman{table}}
\renewcommand{\bibnumfmt}[1]{[S#1]}
\renewcommand{\citenumfont}[1]{S#1}

This Supplemental Material consists of three sections. In Section I~\ref{SMSec:Polarization operator} we evaluate the polarization operator at finite temperature. In Section II~\ref{SMSec:Coulomb} we calculate the imaginary part of the fermionic self-energy with the Coulomb interaction. Finally, in Section III~\ref{SMSec:Hubbard} we perform the same calculation for the case of repulsive Hubbard  interaction. The results of the Supplemental Material are summarized in Eqs.~(3), (8), (10) and Table I of the main text.

\section{I. Polarization operator at finite temperature \label{SMSec:Polarization operator}}

In this Section we calculate the polarization operator at finite temperature. After analytical continuation on real frequencies, it is given by the expression \cite{SMAGD}:

\begin{align}
\text{Re} \, \Pi^R (\omega,\bq) &= -\frac1{(2\pi)^{4}}\int d^3\bp \, d\ve \, \tanh\frac{\ve}{2T} \text{Tr} \, \left[ G''(\ve, \bp) G'(\ve - \omega,\bp - \bq) + G''(\ve, \bp - \bq) G'(\ve+\omega, \bp)   \right], \nonumber \\
\text{Im} \, \Pi^R ( \omega, \bq) &= \frac1{(2\pi)^{4}} \int d^3\bp \, d\ve \, \left( \tanh\frac{\ve}{2T} - \tanh\frac{\ve - \omega}{2T}   \right) \text{Tr} \,  G'' (\ve, \bp) G'' (\ve - \omega, \bp - \bq) ,
\end{align}
where the (bare) fermionic Green's function has form

\be
G^R(\ve, \bp) = \frac12 \left\{ \frac{I + \frac{ \bp \cdot \boldsymbol{\sigma}}{p}}{\ve - \vf p + i0^+}  + \frac{I - \frac{ \bp \cdot \boldsymbol{\sigma}}{p}}{\ve + \vf p + i0^+} \right\},
\ee
and we defined $G' = (G^R + G^A)/2,$ $G'' = (G^R - G^A)/2i.$ The vector of Pauli matrices $\boldsymbol{\sigma}$ obey the standard relation $\text{Tr}\, \sigma^i \sigma^j = 2 \delta^{ij}.$

To evaluate the above integrals, we follow the approach used in Ref.~\onlinecite{SMMirlin2011}. In particular, we use elliptical coordinates defined as $\xi = (p+|\bp - \bq|)/q$ and $\eta = (p-|\bp - \bq|)/q.$ The expression for polarization operator then takes form

\begin{align}
\text{Re} \, \Pi^R (\Omega, Q) &= \frac{T^2}{\vf^3} \frac{Q^2}{2\pi^2}\int_1^{\tilde \Lambda/Q} d\xi \int_0^1 d\eta \frac1{\cosh Q \xi + \cosh Q \eta}\left[\frac{\eta(\xi^2-1)}{\eta^2-\beta^2} \sinh Q\eta + \frac{\xi(1-\eta^2)}{\xi^2-\beta^2} \sinh Q\xi  \right], \\ \text{Im} \, \Pi^R ( \Omega, Q) &= \frac{T^2}{\vf^3} \frac{Q^2 \sinh\Omega}{4\pi}\left[ \Theta(Q^2 - \Omega^2) \int_1^\infty d\xi \frac{\xi^2 - 1}{\cosh\Omega + \cosh\xi Q} + \Theta(\Omega^2 - Q^2)\int_0^1 d\eta \frac{1-\eta^2}{\cosh\Omega + \cosh \eta Q}       \right],
\end{align}
where $\Lambda$ is an ultra-violet momentum cutoff beyond which the spectrum cannot be approximated as linear anymore, and we introduced dimensionless variables $Q = \vf q/2T,$ $\Omega = \omega/2T,$ $\beta = \Omega / Q$, $\tilde \Lambda = \vf \Lambda/2T.$  $\Theta(x)$ is the Heaviside step function. After straightforward calculation, we find in different limiting cases:

\begin{align}
&\text{Re} \, \Pi^R (\Omega, Q) =\frac{T^2}{\vf^3} \left\{ \begin{array}{ccc} \frac16\left( 1 - \frac{|\Omega|}{2Q}\ln\frac{|\Omega|+Q}{||\Omega|-Q|}  \right) + \frac{Q^2}{3\pi^2}\ln \frac{\tilde \Lambda}{\max \{ 1, |\Omega| \}}, & Q\ll 1, & \qquad \text{regions 1,2} \\ \frac{Q^2}{3\pi^2}\ln \frac{\tilde \Lambda}{\sqrt{|\Omega^2 - Q^2|}}, & Q\gg1, &  \qquad \text{regions 3,4}    \end{array}      \right. \nonumber \\
&\text{Im} \, \Pi^R (\Omega, Q) = \frac{T^2}{\vf^3} \left\{  \begin{array}{ccc} \frac{Q^2}{6\pi}\tanh\frac{\Omega}2, & Q\ll1, \, Q<|\Omega|, & \qquad \text{region 1}\\ \frac{\pi}{12}\frac{\Omega}Q,  & Q\ll1, \, Q>|\Omega|, & \qquad \text{region 2} \\ \frac{Q^2}{6\pi}\text{sign} \Omega, & Q\gg1, \, Q<|\Omega|, & \qquad \text{region 3} \\ \frac1{\pi} \sinh\Omega e^{-Q}, & Q\gg1, \, Q>|\Omega|, & \qquad \text{region 4} \end{array}     \right. \label{SMPi}
\end{align}
Under the assumption we made in this paper, $\alpha N \ln \tilde \Lambda \ll 1$ (or $\lambda N T^2 \vf^{-3} \ln \tilde \Lambda \ll 1$ for the Hubbard interaction), the second term in the expression for $\text{Re} \, \Pi^R (\Omega,Q)$ at $Q\ll 1$, $(Q^2/3\pi^2)\ln \tilde \Lambda,$ can be neglected for our further purposes.

As we will demonstrate below, asymptotic expression~(\ref{SMPi}) is sufficient for calculating fermionic self-energy with correct numerical prefactors in a wide energy range. The only exception is an exponentially small region where main contribution comes from energy and momentum transfer of order of temperature. In this regime, the approximate form of polarization operator~(\ref{SMPi}) is not accurate, and, hence, the correct numerical coefficient cannot be obtained.

\section{II. Self-energy due to the Coulomb interaction \label{SMSec:Coulomb}}

\begin{figure*}
\includegraphics[width=0.75\textwidth]{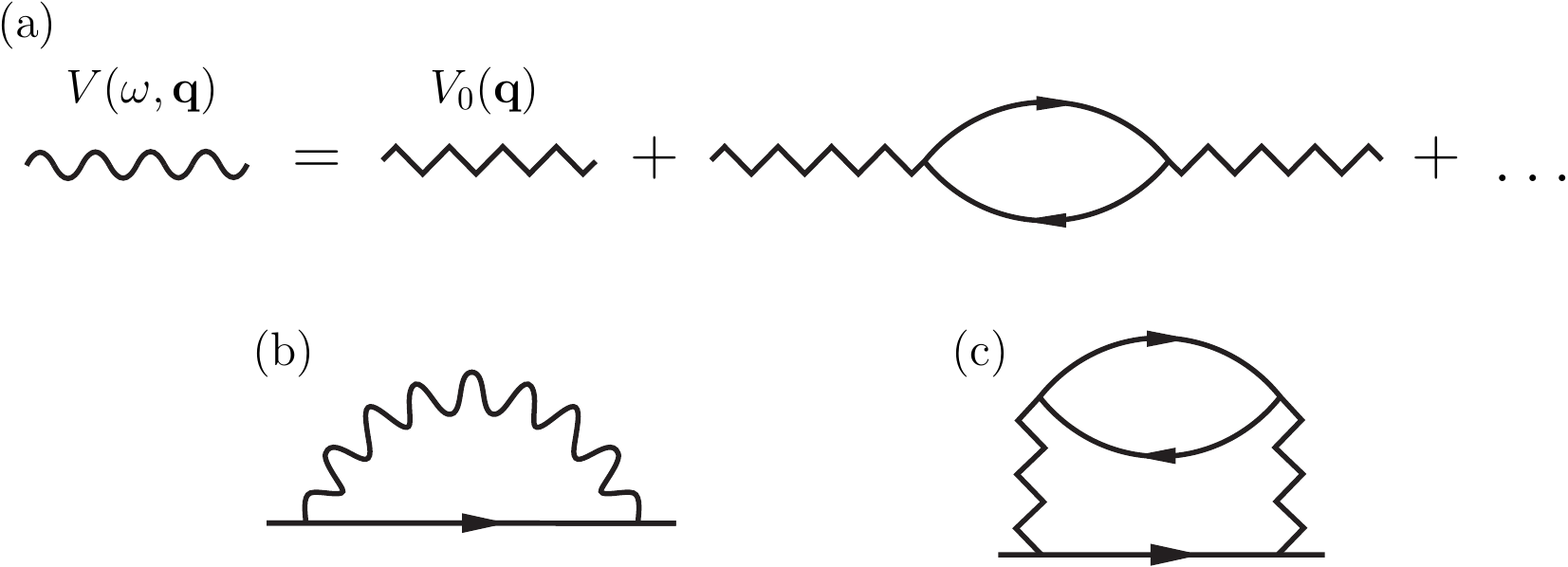}
\caption{(a) The effective interaction $V(\omega, \bq)$ (wavy line) within the RPA, which is justified in the large-$N$ limit. Zigzag line $V_0(\bq)$ corresponds to the bare interaction. (b) One-loop diagram for the fermionic self-energy. (c) The second-order perturbative contribution to the fermionic self-energy. Non-RPA second-order diagram can be neglected within the large-$N$ approximation. }
\label{SMFig}
\end{figure*}

In this Section we present some details of the calculation of the imaginary part of the self-energy due to the Coulomb interaction. In the one-loop approximation, it is given by the diagram shown in Fig.~\ref{SMFig} (b), and its analytical expression reads as~\cite{SMAGD}

\be
\text{Im} \, \Sigma^R(\ve, \bk) = - \frac{1}{(2\pi)^{4}}\int d^3 \bq \, \int d\omega\,  G''( \ve - \omega, \bk -\bq) V'' (\omega,  \bq) \left(\coth\frac{\omega}{2T} + \tanh\frac{\ve - \omega}{2T} \right), \label{SMImSigma1}
\ee
where $G'' = (G^R - G^A)/2i,$ $V'' = (V^R - V^A)/2i.$ Throughout this paper, we assume that the effective interaction  $V(\omega ,\bq)$ is given by the random phase approximation (RPA) series, see Fig.~\ref{SMFig} (a):

\be
V^R(\omega, \bq) = \frac{V_0(\bq)}{1+V_0(\bq)N \Pi^R(\omega, \bq)}, \label{SMEq:V}
\ee
where $V_0(\bq)$ is the bare interaction. The RPA is justified if the number of Weyl/Dirac nodes $N$ is large, which we assume in this paper.

As was discussed in the main text, the self-energy is generally a matrix and can be parameterized as $\Sigma(\ve,\bk) = \Sigma_{\ve} I + \Sigma_v \boldsymbol{\sigma} \cdot \hat \bk.$ Since we are interested in $\Sigma_{\ve}$, one has $G''(\ve,\bp) \to -(\pi/2) [\delta(\ve - \vf p) + \delta(\ve + \vf p)],$ which leads to

\begin{multline}
\text{Im} \, \Sigma^R_{\ve}(\ve,\bk) = \frac14 \sum_{i = \pm} \int \frac{d^3 \bq}{(2\pi)^3} \text{Im} V^R(\omega_i, \bq) \left(\coth\frac{\omega_i}{2T} + \tanh\frac{\ve - \omega_i}{2T}  \right) = \\ = -\frac{1}{16 \pi^2}\sum_{i= \pm} \int_{-1}^1 dt \int_0^{\Lambda} q^2 \, dq \frac{N V_0^2(\bq) \text{Im} \Pi^R(\omega_i,\bq)}{\left[1 + N V_0(\bq) \text{Re} \Pi^R(\omega_i,\bq)   \right]^2 + \left[ N V_0(\bq) \text{Im} \Pi^R(\omega_i,\bq) \right]^2} \left(\coth\frac{\omega_i}{2T} + \tanh\frac{\ve - \omega_i}{2T}  \right) , \label{SMImSigma2}
\end{multline}
with $\omega_\pm \equiv \ve \pm \vf |\bk - \bq| = \ve \pm \vf \sqrt{k^2 + q^2 - 2 k q t},$ $t= \cos \theta_{\bk \bq},$ and $\theta_{\bk \bq}$ is an angle between vectors $\bk$ and $\bq$. In dimensionless variables, the self-energy takes form

\be
\text{Im} \, \Sigma^A_{\ve}(x,y) =  \frac{1}{2\pi^2}\frac{T^3}{\vf^3}\sum_{i= \pm} \int_{-1}^1 dt \int_0^{\tilde\Lambda} Q^2 \, dQ \frac{N V_0^2(Q) \text{Im} \Pi^R(\Omega_i,Q)\left[\coth\Omega_i + \tanh(x- \Omega_i) \right]}{\left[1 + N V_0(Q) \text{Re} \Pi^R(\Omega_i,Q)   \right]^2 + \left[ N V_0(Q) \text{Im} \Pi^R(\Omega_i,Q) \right]^2}, \label{SMImSigma3}
\ee
where $\Omega_{\pm} = x \pm \sqrt{y^2 + Q^2 - 2 y Q t},$ $ x = \ve/2T,$ $y = \vf k /2T,$ and we focus on the advanced Green's function hereafter to get rid of an extra "$-$" sign. This expression is very general and will serve as a starting point for both the Coulomb and the Hubbard interactions.

In the case of the Coulomb interaction, $V_0(Q)$ is given by

\be
V_0(Q) = \pi \alpha \frac{\vf^3}{T^2}\frac1{Q^2},
\ee
where $\alpha = e^2 / \epsilon \vf$ is an effective fine-structure constant.

Now we have all ingredients to calculate $\text{Im} \, \Sigma$ in different limiting cases. In this paper we focus on a weak-coupling limit. Moreover, to simplify some of the final expressions, we assume that $\alpha N \ln \tL \ll 1.$

\subsection{A. Calculation of $\text{Im} \, \Sigma_{\ve} $ on the mass shell (scattering rate)}

On the mass shell, fermionic energy and momentum are related as $y = x$ (or $\vf k = \ve$), and we consider positive energies for definiteness. Thus, we have $\Omega_{\pm} = x \pm \sqrt{x^2 + Q^2 - 2 x Q t}$. There are three main energy domains, where the integration can be performed and the expression for $\text{Im} \, \Sigma$ can be obtained analytically.

\subsubsection{1. Regime $x \gg \sqrt{\alpha N}$}

In the regime $x \gg \sqrt{\alpha N}$ the main contribution comes from region 2, see Eq.~(\ref{SMPi}), defined as $Q\ll 1,$ $|\Omega|<Q.$ As we demonstrate below, important momenta are $Q\sim \sqrt{\alpha N} \ll \min\{x, 1\}$. The corresponding contribution to the scattering rate then equals (only $\Omega_- \approx Q t \ll 1$ contributes)

\begin{align}
\frac1{\tau(x)} \equiv - 2 \text{Im}\, \Sigma^R_{\ve}(x,x) &=\frac{\pi}{12} \alpha^2 N T \int_{-1}^1 dt \int\limits_0^{\sim\min\{ x, 1 \}} \frac{Q \Omega_-\left[\coth\Omega_-+ \tanh(x-\Omega_-)\right]dQ}{\left[Q^2 + \frac{\pi \alpha N}6\left( 1 - \frac{|\Omega_-|}{2Q} \ln\frac{Q+|\Omega_-|}{Q-|\Omega_-|}  \right)  \right]^2    + \left[ \frac{\alpha N \pi^2}{12}\frac{\Omega_-}Q \right]^2} \approx \nonumber \\ &\approx  \frac{\pi}{12} \alpha^2 N T \int_{0}^1 dt \int_0^{\infty} \frac{2 Q dQ}{\left[Q^2 + \frac{\pi \alpha N}6\left( 1 - \frac{t}{2} \ln\frac{1+t}{1-t}  \right)  \right]^2    + \left[ \frac{\alpha N \pi^2}{12}t \right]^2} = \nonumber \\ &= \alpha T \int_0^1 \frac{dt}{2\pi t}\left\{ \pi - 2\arctan\left[\frac2{\pi t}\left(1- \frac{t}2\ln \frac{1+t}{1-t}  \right)   \right]  \right\} \approx 0.549 \alpha T.
\end{align}
We see that, indeed, the main contribution comes from $Q \sim \sqrt{\alpha N} \ll \min\{ x,1\}$, hence, we changed the upper limit of integration over $Q$ from $\sim \min\{ x,1\}$ to $\infty$.
This result is non-perturbative and requires the RPA summation.

\subsubsection{2. Regime $\exp\left(-\frac{c_2}{\alpha N}  \right) \ll x \ll \sqrt{\alpha N}$}

In this regime, $\exp\left(-c_2/\alpha N  \right) \ll x \ll \sqrt{\alpha N},$ where $c_2$ is some numerical constant of order 1, the main contribution comes from regions 1 and 2, implying that $Q\ll1$. In region 1, i.e., $Q\ll1,$ $Q<|\Omega|$, only $\Omega_+$ contributes, since $|\Omega_-|<Q$. Moreover, as we show below, the important momenta are $Q \sim \sqrt{ \alpha N \ln \frac{\sqrt{\alpha N}}{x}} \gg x$, so we can write

\begin{align}
\left( \frac1{\tau(x)} \right)_1 &= \frac{\alpha^2 N T}{6\pi} \int_{-1}^1 dt \int_{\sim x}^{\sim 1}dQ \frac{\tanh(\Omega_+/2)\left[ \coth \Omega_+ + \tanh(x-\Omega_+) \right]}{\left[ 1+ \frac{\pi \alpha N}{6 Q^2}\left( 1 - \frac{|\Omega_+|}{2Q}\ln \frac{|\Omega_+|+Q}{|\Omega_+|-Q} \right) \right]^2 + \left[ \frac{\alpha N}6 \tanh\frac{\Omega_+}2 \right]^2} \approx \nonumber \\ &\approx \frac{\alpha^2 N T}{12\pi} \int_{-1}^1 dt \int_{\sim x}^{\sim 1} \frac{Q^4 dQ}{\left[ Q^2 + \frac{\pi \alpha N}6 \left( 1- \frac12 \ln \frac{2Q}{x(1-t)}  \right) \right]^2 + \left[ \frac{\alpha N Q^3}{12}  \right]^2},
\end{align}
where we used the fact that $\Omega_+ \approx Q$ for $Q\gg x$. We see that, indeed, the main contribution comes from the vicinity of $Q_0$ determined by the equation

\be
Q_0^2 = \frac{\pi \alpha N}6 \left(  \frac12 \ln \frac{2Q_0}{x(1-t)} -1 \right). \label{SMEq:Q0}
\ee
With the logarithmic accuracy, the solution is given by $Q_0 \approx \sqrt{ \frac{\pi \alpha N}{12} \ln \frac{\sqrt{\alpha N}}{x}}$ and satisfies $x \ll Q_0 \ll 1.$ Hence, assuming also that $\ln(\sqrt{\alpha N}/x) \gg 1$, we obtain

\be
\left( \frac1{\tau(x)} \right)_1 \approx \frac{\alpha^2 N T}{12\pi} \int_{-1}^1 dt \int_{\sim x}^{\sim 1}\frac{Q_0^4 dQ}{(Q^2 - Q_0^2)^2 + \left( \frac{\alpha N Q_0^3}{12} \right)^2} \approx \frac{\alpha^2 N T Q_0^2}{24 \pi}\int_{-\infty}^{\infty} \frac{d Q}{Q^2 + \left( \frac{\alpha N Q_0^2}{24} \right)^2} = \alpha T.
\ee

Analogously, we find the contribution from  region 2, $|\Omega|<Q$, which only exists for the term with $\Omega_-\approx -Q$ (since $|\Omega_+|>Q$):

\begin{align}
\left( \frac1{\tau(x)} \right)_2 &= \frac{\pi \alpha^2 N T}{12} \int_{-1}^1 dt \int_{\sim x}^{\sim 1}\frac{\Omega_- dQ}{Q^3} \frac{\left[ \coth \Omega_- + \tanh(x-\Omega_-) \right]}{\left[ 1+ \frac{\pi \alpha N}{6 Q^2}\left( 1 - \frac{|\Omega_-|}{2Q}\ln \frac{Q+|\Omega_-|}{Q - |\Omega_-|} \right) \right]^2 + \left[ \frac{\pi^2\alpha N \Omega_-}{12 Q^3}  \right]^2} \approx \nonumber \\ &\approx \frac{\pi \alpha^2 NT Q_0}6 \int_{\sim x}^{\sim 1} \frac{dQ}{(Q^2-Q_0^2)^2+\left( \frac{\pi^2 \alpha N}{12} \right)^2} \approx \frac{\pi \alpha^2 NT }{24 Q_0} \int_{-\infty}^{\infty} \frac{dQ}{Q^2+\left( \frac{\pi^2 \alpha N}{24 Q_0} \right)^2} = \alpha T,
\end{align}
where $Q_0$ is determined by the same expression~(\ref{SMEq:Q0}). Here, again, we assumed that $\ln(\sqrt{\alpha N}/x) \gg 1$, while the corrections to the above expression are of order $1/\ln(\sqrt{\alpha N}/x)$.

Summing up the two contributions, we find

\be
 \frac1{\tau(x)} = \left( \frac1{\tau(x)} \right)_1 + \left( \frac1{\tau(x)} \right)_2  = 2\alpha T,  \qquad \exp\left(-\frac{c_2}{\alpha N } \right) \ll x \ll \sqrt{\alpha N}.
\ee

One can easily show that these contributions remain the same for any energy and momentum $\ve$ and $\bk$ satisfying the condition $T\exp\left(-\frac{c_2}{\alpha N}  \right) \ll \max\{ \ve, \vf k  \} \ll T\sqrt{\alpha N},$ independently of the ratio between $\ve$ and $\vf k$. Hence, we conclude that

\be
\text{Im} \, \Sigma^A_\ve(\ve, k) \approx \alpha T, \qquad T\exp\left(-\frac{c_2}{\alpha N } \right) \ll \max\{\ve, \vf k  \} \ll T \sqrt{\alpha N}.
\ee
A word of caution is needed here. We assumed, but did not check it in general case, that there are no significant contributions from any regions other than the vicinity of $Q_0 \approx \sqrt{ \frac{\pi \alpha N}{12} \ln \frac{T\sqrt{\alpha N}}{\max\{ \ve, \vf k \}}}$. We did check this statement, however, for the cases we focused on in this paper, namely, for zero energy, momentum, and on the mass shell.

\subsubsection{3. Regime $x \ll \exp\left(-\frac{c_2}{\alpha N}  \right)$}

For the exponentially small energies, $x \ll \exp\left(-\frac{c_2}{\alpha N}  \right)$, all regions contribute to the scattering rate. The main contribution comes from $|\Omega| \sim Q \sim 1$, where our approximation for the polarization operator Eq.~(\ref{SMPi}) is not accurate, and, as a result, the scattering rate cannot be found with the correct numerical prefactor. The functional dependence, however, can still be extracted simply by extrapolating expression~(\ref{SMPi}) to the region $Q\sim1.$ To demonstrate it, we consider region 3, $Q\gg1$, $|\Omega|>Q$. One can estimate the contribution from this region as

\be
\frac1{\tau(x)} \sim \alpha^2 N T \int_{-1}^1 dt \int_{\sim 1}^{\infty} \frac{d Q}{Q^2}
\frac{\coth\Omega_+ + \tanh(x-\Omega_+)}{\left( 1+ \frac{\alpha N }{3\pi}\ln\frac{\tilde \Lambda}{\sqrt{\Omega_+^2 - Q^2}}  \right)^2 + \left(\frac{\alpha N}6  \right)^2} \sim \alpha^2 NT \int_{\sim 1} \frac{dQ}{Q^2}\frac1{[\alpha N \ln(xQ)]^2} \sim \frac{T}{N(\ln x)^2}, \label{SMEq:expsmallenergies}
\ee
where we exploited the fact that $\Omega_+^2 - Q^2 \approx 2x Q(1-t)$ and $\alpha N \ln(1/x) \gg 1$. The contributions from all other regions can be estimated analogously, leading to the same answer. Important, Eq.~(\ref{SMEq:expsmallenergies}) results from the logarithmical factor in the {\it real} part of the polarization operator.

The same behavior of $\text{Im} \, \Sigma (\ve, \bk)$ is expected for all exponentially small energies/momenta,  $\max\{\ve, \vf k  \} \ll T\exp\left(-c_2/\alpha N  \right),$ not only on the mass shell.

\subsubsection{4. Plasmon-enhanced peak at $x= \Omega_{\text{pl}}/2$ \label{SMSec:plasmonpeakonmassshell}}

When frequency and momentum of the effective interaction $V(\Omega, Q)$ satisfy the plasmon dispersion relation, $\Omega = \Omega_{\text{pl}} + 3 Q^2/10 \Omega_{\text{pl}}$ with $\Omega_{\text{pl}} = \sqrt{\pi \alpha N/18}$ (and assuming that $Q\lesssim \Omega_{\text{pl}}$), the scattering rate exhibits significant enhancement due to thermal plasmons. This condition can only be satisfied in region 1, $Q\ll1$ with $Q<|\Omega|$, provided $x= \Omega_{\text{pl}}/2$. The expression for scattering rate then takes form


\be
\frac{1}{\tau(x = \Omega_{\text{pl}}/2)} = \frac{\alpha^2 NT}{12 \pi}\int_{-1}^{1} dt \int_0^{\sim 1} \frac{dQ}{\left[ 1+ \frac{\pi \alpha N}{6Q^2}\left( 1 - \frac{\Omega_+}{2Q}\ln \frac{\Omega_+ + Q}{\Omega_+ -Q}   \right)   \right]^2 + \left[ \frac{\alpha N \Omega_+}{12}\right]^2}.
\ee
Expanding the expression for $\Omega_+$ at $Q \ll x$, we find

\be
1+ \frac{\pi \alpha N}{6Q^2}\left( 1 - \frac{\Omega_+}{2Q}\ln \frac{\Omega_+ + Q}{\Omega_+ -Q}   \right) = \frac Qx \left( \frac{7Q}{20x} -t \right),
\ee
resulting in

\be
\frac{1}{\tau\left(x = \frac{\Omega_{\text{pl}}}2\right)} = \frac{\alpha^2 NT x^2}{12 \pi} \int_0^{\sim x} \frac{dQ}{Q^2}\int_{-1}^{1} \frac{dt}{\left(t -\frac{7Q}{20x}  \right)^2 + \left( \frac{\alpha N x^2}{6Q}\right)^2}\approx \frac{\alpha T}{2}\int_{\sim \alpha N x^2}^{\sim x} \frac{dQ}Q= \frac{\alpha T}2 \ln\frac1{\alpha N x} \approx \frac{3\alpha T}4 \ln \frac1{\alpha N}.
\ee
We see that the main contribution comes from $t = 7 Q /20x,$ which, plugged into the expression for frequency, leads exactly to the plasmon dispersion, $\Omega = x + \sqrt{x^2 + Q^2 - 2 Q x t} \approx 2 x + 3Q^2/20 x = \Omega_{\text{pl}} + 3 Q^2/10 \Omega_{\text{pl}}.$ Momenta that contribute belong to the region $\Omega^4_{\text{pl}} \lesssim Q \lesssim \Omega_{\text{pl}}$, thus justifying our original assumption.

\subsection{B. Calculation of $\text{Im} \, \Sigma_{\ve} (\ve=0, k)$}

The calculation in case of zero energy, $\ve=0$, is very similar to that on the mass shell. The imaginary part of self-energy is given by the same expression~(\ref{SMImSigma3}), but now with $x=0$ and  $\Omega_\pm = \pm \Omega \equiv \pm \sqrt{Q^2 + y^2 -2 Q y t}$ (here $y = \vf k /2T$), leading to

\be
\text{Im} \, \Sigma^A_{\ve}(0,y) = -\frac{2T^3}{\pi^2 \vf^3} \int_{-1}^1 dt \int_0^{\tilde \Lambda} Q^2 dQ \frac{\text{Im} V^R (\Omega, Q)}{\sinh 2\Omega}. \label{SMEq:zeroenergy}
\ee

\subsubsection{1. Regime $y \gg 1$}

In the regime $y \gg 1$ the main contribution comes from region 3, $Q\gg1$, $|\Omega|>Q$. The condition $|\Omega|>Q$ implies that $Q t < y/2.$ We find

\be
\text{Im} \, \Sigma^A_{\ve}(0,y) \approx \frac{\alpha^2 NT}{3\pi} \int_{-1}^1 dt \int_{\sim 1}^{\infty} dQ\frac{\Theta(y - 2 Q t)}{\sinh 2\Omega} \approx \frac{2\alpha^2 NT}{3\pi} \int_{-1}^1 dt \int_{\sim 1}^{\infty} dQ\Theta(y - 2 Q t) e^{-2\Omega}.
\ee
Thus, the most significant contribution comes from the region where $\Omega = \sqrt{Q^2+y^2-2Q y t}$ is at minimum, resulting in $Q\approx y/2,$ $t\approx1.$ Hence, changing variables of integration according to $Q = y(1+\xi)/2$, $t = 1- \eta,$ we find that $\Omega \approx y(1-\xi + 2 \eta)/2$ and the condition $2 Q t < y$ is equivalent to $\xi < \eta.$ This leads to

\be
\text{Im} \, \Sigma^A_{\ve}(0,y) \approx \frac{\alpha^2 N T y}{3\pi}\int_0^\infty d\eta \int_{-\infty}^\eta d\xi e^{-y + y(\xi - 2\eta)} = \frac{\alpha^2 N T y}{3\pi}e^{-y}\int_0^\infty d\eta e^{-y \eta} \int_0^{\infty} d\xi e^{-y \xi } = \frac{\alpha^2 N T}{3\pi} \frac{e^{-y}}y.
\ee
We see that, unlike the case on the mass shell, this result is within the second-order perturbation theory and can be obtained, in principle, without the RPA summation.

\subsubsection{2. Regime $\sqrt{\alpha N} \ll y \ll 1$}

In this regime the main contribution comes from region 2, $Q\ll1$ with $|\Omega|<Q,$ which implies that $2 Q t>y.$ This leads to

\be
\text{Im} \, \Sigma^A_{\ve}(0,y) \approx \frac{\pi \alpha^2 NT}{6} \int_{y/2}^{\sim 1} \frac{dQ}{Q^3}\int_{y/2Q}^1 dt\frac{\Omega}{\sinh2\Omega}\frac1{\left[ 1+\frac{\pi \alpha N}{6 Q^2}\left(1-\frac{\Omega}{2Q}\ln \frac{Q+\Omega}{Q-\Omega}  \right)\right]^2 + \left[ \frac{\pi^2 \alpha N \Omega}{12 Q^3} \right]^2}.
\ee
More specifically, important frequencies and momenta are $\Omega\sim Q \sim y \gg \sqrt{\alpha N},$ which results in

\be
\text{Im} \, \Sigma^A_{\ve}(0,y) \approx \frac{\pi \alpha^2 NT}{12} \int_{y/2}^{\infty} \frac{dQ}{Q^3}\left( 1 - \frac{y}{2Q}  \right) = \frac{\pi \alpha^2 N T}{18 y^2}, \qquad \sqrt{\alpha N} \ll y \ll 1,
\ee
where we extended the limit of integration over $Q$ from $\sim 1$ to $\infty$. This result is, again, within the second-order perturbation theory.

\subsubsection{3. Regime $y \ll \sqrt{\alpha N}$}

The calculation in this region is absolutely analogous to the calculation on the mass shell, leading to the same result:

\be
\text{Im} \, \Sigma^A_{\ve}(0,y) = \left\{ \begin{array}{cc} \sim \frac{T}{N\ln^2 y}, & y\ll \exp\left( -\frac{c_2}{\alpha N}  \right) \\ \alpha T, &  \exp\left( -\frac{c_2}{\alpha N}  \right) \ll y \ll \sqrt{\alpha N}    \end{array}      \right.
\ee
This result is non-perturbative and requires the RPA summation.

\subsubsection{4. Plasmon-enhanced peak at $y= \Omega_{\text{pl}}$}

The calculation of plasmon peak at zero external energy is similar to the on-mass-shell case, with the only difference that the peak position now is at $y= \Omega_{\text{pl}}$. Repeating all the steps of Sec. II A 4~\ref{SMSec:plasmonpeakonmassshell}, we find

\be
1+ \frac{\pi \alpha N}{6Q^2}\left( 1 - \frac{\Omega}{2Q}\ln \frac{\Omega + Q}{\Omega -Q}   \right) = \frac {2Q}y \left( \frac{Q}{5y} -t \right),
\ee
leading to

\be
\text{Im} \, \Sigma^A_\ve(0,y=\Omega_{\text{pl}}) = \frac{\alpha^2 NT y^2}{48 \pi} \int_0^{\sim y} \frac{dQ}{Q^2}\int_{-1}^{1} \frac{dt}{\left(t -\frac{Q}{5y}  \right)^2 + \left( \frac{\alpha N y^2}{24Q}\right)^2}\approx \frac{\alpha T}{2}\int_{\sim \alpha N x^2}^{\sim x} \frac{dQ}Q= \frac{\alpha T}2 \ln\frac1{\alpha N y} \approx \frac{3\alpha T}4 \ln \frac1{\alpha N}.
\ee
Plugging $t = Q /5y$ back into the expression for bosonic frequency, we find again $\Omega = \sqrt{y^2 + Q^2 - 2 Q x t} \approx  y + 3Q^2/10 y = \Omega_{\text{pl}} + 3 Q^2/10 \Omega_{\text{pl}},$ which is the plasmon dispersion. As before, momenta that contribute belong to the region $\Omega^4_{\text{pl}} \lesssim Q \lesssim \Omega_{\text{pl}}.$

\subsection{C. Calculation of $\text{Im} \, \Sigma_{\ve} (\ve, k=0)$}

In case of zero external momentum, $\bk = 0$, the integration over $t$ in Eq.~(\ref{SMImSigma3}) can be performed explicitly, resulting in

\be
\text{Im} \, \Sigma^A_{\ve}(x,0) = -\frac{T^3}{\pi^2 \vf^3}\sum_{i=\pm}\int Q^2 dQ \, \text{Im} \, V^R (\Omega_i,Q) \left[ \coth\Omega_i + \tanh(x-\Omega_i) \right], \label{SMEq:zeromomentum}
\ee
with $\Omega_{\pm} = x \pm Q.$

\subsubsection{1. Regime $x\gg1$}

The most significant contribution comes from the term with $\Omega_-$ in region 3, $Q \gg 1$ with $|\Omega_-|>Q,$ which implies that $2Q<x.$ The evaluation of integral in this case is straightforward:

\be
\text{Im} \, \Sigma^A_{\ve}(x,0) \approx \frac{\alpha^2 N T}{6\pi} \int_{\sim 1}^{x/2} dQ \left[ \coth(x-Q) + \tanh(Q)  \right] \approx \frac{\alpha^2 N T x}{6 \pi}.
\ee
The important momenta are $Q\le x/2.$

\subsubsection{2. Regime $\sqrt{\alpha N} \ll x \ll1$ }

The main contribution in this regime comes from the term with $\Omega_-$ in region 2, $Q\ll1$ with $|\Omega_-|<Q,$ implying that $x< 2Q.$ The self-energy then equals

\be
\text{Im} \, \Sigma^A_{\ve} (x,0) \approx \frac{\pi \alpha^2 NT}{12}\int_{x/2}^{\sim 1}\frac{dQ}{Q^3} \approx \frac{\pi \alpha^2 N T}{6 x^2}.
\ee
The leading contribution comes from $\Omega\sim Q \sim x.$

\subsubsection{3. Regime $x \ll \sqrt{\alpha N}$}

The calculation in this region is absolutely equivalent to the calculation on the mass shell, leading to the same result:

\be
\text{Im} \, \Sigma^A_{\ve}(x,0) = \left\{ \begin{array}{cc} \sim \frac{T}{N\ln^2 x}, & x\ll \exp\left( -\frac{c_2}{\alpha N}  \right) \\ \alpha T, &  \exp\left( -\frac{c_2}{\alpha N}  \right) \ll x \ll \sqrt{\alpha N}    \end{array}      \right.
\ee

\subsubsection{4. The absence of plasmon-enhanced peak}

At zero external momentum, $\bk = 0$, the bosonic frequencies are given by $\Omega_{\pm} = x \pm Q$. This implies that the plasmon dispersion relation can never be satisfied, independently of $x$. As a result, there is no plasmon peak in $\text{Im} \, \Sigma_{\ve}(x,0)$.

To demonstrate it explicitly, we repeat the same calculation as in previous sections. First, consider the contribution from $\Omega_+ = x + Q$. For $Q \ll \Omega_+$, we find at $x = \Omega_{\text{pl}}$ (which is the best candidate for plasmon resonance)

\be
1+ \frac{\pi \alpha N}{6Q^2}\left( 1 - \frac{\Omega_+}{2Q}\ln \frac{\Omega_+ + Q}{\Omega_+ -Q}   \right) \approx \frac {2Q}x.
\ee
This yields the contribution to self-energy

\be
\left[\text{Im} \, \Sigma^A_{\ve} (x = \Omega_{\text{pl}},0)\right]_1 \approx \frac{\alpha^2 N T}{12\pi}\int_0^{\sim x} \frac{dQ}{\left(\frac{2Q}{x}\right)^2+\left( \frac{\alpha N x}{12} \right)^2} \approx \frac{\alpha T}4.
\ee
The same contribution is obtained from the term with $\Omega_-$, resulting in

\be
\text{Im} \, \Sigma^A_{\ve} (x = \Omega_{\text{pl}},0) = \frac{\alpha T}2.
\ee

\section{III. Self-energy due to the Hubbard interaction \label{SMSec:Hubbard}}

In the case of the Hubbard interaction we neglect the inter-nodal scattering, so the imaginary part of self-energy is determined again by Eq.~(\ref{SMImSigma3}) with $V_0(\bq) = \lambda>0.$ We also assume that the effective interaction is given by the 'bubble' series Eq.~(\ref{SMEq:V}), diagrammatically shown in Fig.~\ref{SMFig} (a), which is justified in the large-$N$ limit. It is convenient to introduce dimensionless coupling constant $\tl \equiv \lambda T^2/\vf^3.$ We emphasize that defined this way $\tl$ is temperature-dependent. The weak-coupling limit we consider in this paper corresponds to $\tl N \ll 1$. To simplify final expressions, we also assume that  $\tl N \ln \tL \ll 1.$

\subsection{A. Regime $\max\{\ve, \vf k  \} \ll T$ ($\max\{x, y \} \ll 1$)}

Before focusing on different limiting cases, we comment on the regime $\max\{\ve, \vf k  \} \ll T$. It can be shown that the self-energy in this regime is determined by bosonic frequencies and momenta of order of temperature ($\Omega \sim Q \sim 1$).  It implies that the numerical prefactors in this regime cannot be determined within the approach we use in this paper. It is straightforward to show, however, that self-energy behaves as

\be
\text{Im} \, \Sigma^A_{\ve}(x,y) = \left\{ \begin{array}{cc} \sim \frac{T}{N} \left( \frac1{\ln( \max\{x,y\})} \right)^2, & x,y\ll \exp\left( - b_2/\tl N  \right) \\ \sim \tl^2  N T, &  \exp\left( - b_2/\tl N \right) \ll \max\{x,y  \} \ll 1    \end{array}      \right.
\ee
where $b_2$ is a numerical coefficient of order 1, and we defined, as before, $x \equiv \ve/2T,$ $y\equiv \vf k /2T$. The evaluation is absolutely analogous to the case of the Coulomb interaction. We see that in the regime $\exp\left( -b_2/\tl N  \right) \ll \max\{x,y  \} \ll 1$ the result is proportional to $\text{Im} \, \Sigma^A_{\ve}(x,y)  \propto \tl^2 N,$ hence, one can expect that the correct numerical prefactor can be extracted by means of the second-order perturbation theory. Indeed, we perform perturbative calculation in Sec. III E~\ref{SMSec:Perturb} and find that

\be
\text{Im} \, \Sigma^A_{\ve}(x,y) = 0.035 \tl^2 N T, \qquad \exp\left( -b_2/ \tl N  \right) \ll \max\{x,y  \} \ll 1.
\ee

\subsection{B. Calculation of $\text{Im} \, \Sigma_{\ve} $ on the mass shell (scattering rate)}

On the mass shell fermionic energy and momentum are related as $y = x$ (or $\vf k = \ve$), leading to $\Omega_{\pm} = x \pm \sqrt{x^2 + Q^2 - 2 x Q t}$.

\subsubsection{1. Regime $1 \ll x \ll\ 1/\sqrt{\tl N \ln \tilde \Lambda}$}

In this regime the leading contribution comes from regions 3 ($Q\gg 1$ and $|\Omega_+|>Q$) and 4 ($Q\gg 1$ and $|\Omega_-|<Q$).

The contribution from region 3 equals

\be
\left(\frac1{\tau(x)} \right)_1 = \frac{\tl^2 N T}{6\pi^3} \int_{-1}^1 dt \int_{\sim 1}^{\infty} dQ  \, Q^4 \left[ \coth \Omega_+ + \tanh(x - \Omega_+) \right],
\ee
where $\Omega_+ = x + \sqrt{x^2 + Q^2 - 2 x Q t}.$ Since $\Omega_+ \gg 1$, one can write

\be
\coth \Omega_+ + \tanh(x - \Omega_+)  \approx \frac2{e^{2\sqrt{x^2 + Q^2 - 2 x Q t}}+1}.
\ee
Hence, the important contribution comes from the vicinity of $Q\approx x, \, t \approx 1$. Introducing new variables $\xi$ and $\eta$ as $Q = x(1+\xi)$ and $t = 1-\eta,$ we find

\be
\left(\frac1{\tau(x)} \right)_1 = \frac{\tl^2 N T x^5}{3 \pi^3} \int_0^{\infty} d\eta \int_{-\infty}^{\infty} d \xi \frac1{e^{2 x \sqrt{2\eta + \xi^2}}+1} = \frac{3 \zeta(3) \tl^2 N T x^2}{24 \pi^3}.
\ee

The contribution from region 4 equals

\be
\left(\frac1{\tau(x)} \right)_2 = \frac{\tl^2 N T}{\pi^3}\int_{-1}^1 dt \int_{\sim 1}^{\infty} dQ \, Q^2 e^{-Q} \sinh \Omega_- \left[ \coth \Omega_- + \tanh(x - \Omega_-) \right],
\ee
with $\Omega_- = x - \sqrt{x^2 + Q^2 - 2 x Q t}.$ The leading contribution comes from $Q\le x$ and $t \approx 1$. Then, changing again $t = 1-\eta$ and using $\Omega_-\approx Q - x Q \eta/(x-Q)$, we find

\be
\left(\frac1{\tau(x)} \right)_2  \approx \frac{\tl^2 N T}{\pi^3}\int_0^{\infty} d \eta \int_0^x d Q \, Q^2 e^{-Q + \Omega_-} \approx \frac{\tl^2 N T}{\pi^3} \int_0^x d Q \, Q^2    \int_0^{\infty} d \eta \, e^{-x Q \eta/(x- Q)} = \frac{\tl^2 N T}{6\pi^3} x^2.
\ee

Adding up these two contributions, we find

\be
\frac1{\tau(x)}=\left(\frac1{\tau(x)} \right)_1 + \left(\frac1{\tau(x)} \right)_2 = \frac{3 \zeta(3) +4 }{24 \pi^3} \tl^2 N T x^2. \label{SMEq:Hubbardmass-shellperturb}
\ee
In contrast to the Coulomb case, where perturbation theory is not applicable on the mass shell at all, Eq.~(\ref{SMEq:Hubbardmass-shellperturb}) can be obtained, in principle, within the perturbative calculation.

\subsubsection{2. Regime $ 1/\sqrt{\tl N \ln \tilde \Lambda} \ll x \ll \tL$}

This regime, obviously, only exists provided $\tl N \tL^2 \ln \tL \gg 1.$ The main contribution comes from region 4, namely, from $1 \ll Q\sim 1/\sqrt{\tl N \ln \tL} \ll x,$ $t\approx 1,$ and $\Omega_-\approx Q.$ Introducing new variable $\eta = 1-t$, we easily find

\be
\frac1{\tau(x)} \approx \frac{\tl^2 N T}{\pi^3}\int_{0}^{\infty} d\eta \int_{\sim 1}^{\sim x} dQ \, Q^2\frac{e^{- \eta Q}}{\left( 1+ \frac{\tl N Q^2}{3\pi^2} \ln \frac{\tL}{\sqrt{Q}}  \right)^2} \approx \frac{\tl^2 N T}{\pi^3}\int_0^{\infty}\frac{Q dQ}{\left(1+\frac{\tl N Q^2}{3\pi^2}\ln (\tL \sqrt[4]{\tl N})\right)^2} \approx \frac{3\tl T}{2\pi \ln  \tL}.
\ee
This result is non-perturbative and can only be obtained within the RPA. Important, logarithmical factor arises due to the real part of polarization operator.

\subsection{C. Calculation of $\text{Im} \, \Sigma_{\ve} (\ve=0, k)$}

In case of zero energy, the fermionic self-energy can be found using Eq.~(\ref{SMEq:zeroenergy}).

\subsubsection{1. Regime $1 \ll y \ll\ 1/\sqrt{\tl N \ln \tilde \Lambda}$}

In this regime the main contribution comes from region 3, $Q\gg1$ and $\Omega = \sqrt{Q^2 + y^2 - 2 Q t} > Q$, which implies that $2 Q t \le y$. Self-energy can then be written as

\be
\text{Im}\,\Sigma^A_{\ve}(0,y) \approx \frac{2\tl^2 N T}{3\pi^3}\int\limits_{-1}^{\min \{ 1, y/2Q \}} dt \int_{\sim 1}^{\infty} dQ \, Q^4 e^{-2\Omega},
\ee
where $y=\vf k/2T.$ The leading contribution comes from $Q\approx y/2$ and $t\approx 1$. Hence, defining new variables according to $Q=y(1+\xi)/2$ and $t=1-\eta,$ we find

\be
\text{Im} \, \Sigma^A_{\ve}(0,y) \approx \frac{2 \tl^2 N T}{3\pi^3}\left( \frac{y}2 \right)^5  e^{-y} \int_0^{\infty} d\eta \int_{-\infty}^{\eta} d\xi e^{y(\xi - 2\eta)} = \frac{\tl^2 N T y^3 e^{-y}}{48 \pi^3}.
\ee

\subsubsection{2. Regime $ 1/\sqrt{\tl N \ln \tilde \Lambda} \ll y \ll \tL$}

In this regime, again, the main contribution comes from region 3, namely, from $Q\approx y/2$ and $t\approx 1$. The self-energy can then be easily evaluated:

\be
\text{Im}\, \Sigma^A_{\ve}(0,y) \approx \frac{2 \tl^2 N T}{3\pi^3}\left( \frac{y}2 \right)^5  e^{-y} \int_0^{\infty} d\eta \int_{-\infty}^{\eta} d\xi \frac{e^{y(\xi - 2\eta)}}{\left(\frac{\tl N y^2}{12 \pi^2} \ln \frac{\tL}{y\sqrt{\eta - \xi}}  \right)^2} \approx \frac{3\pi T}{N} \frac{e^{-y}}{y \ln^2\frac{\tL}{\sqrt{y}}}.
\ee
This result is asymptotically correct provided $\ln (\tL/\sqrt{y}) \gg1,$ since the subleading corrections have extra powers of $1/\ln (\tL/\sqrt{y} ).$

\subsection{D. Calculation of $\text{Im} \, \Sigma_{\ve} (\ve, k=0)$}

At zero external momentum, fermionic self-energy can be found using Eq.~(\ref{SMEq:zeromomentum}).

\subsubsection{1. Regime $1 \ll x \ll\ 1/\sqrt{\tl N \ln \tilde \Lambda}$}

Main contribution in this regime comes from the term with $\Omega_-= x - Q$ in region 3. In this region, $x-\Omega_- = Q\gg1$ and $\Omega_- \ge x/2 \gg 1$ (since the condition $Q<|\Omega_-|$ leads to $Q < x/2$). Consequently, $\coth \Omega_- + \tanh(x-\Omega_-) \approx 2,$ and self-energy can be easily evaluated:

\be
\text{Im} \, \Sigma^A_{\ve}(x,0) = \frac{\tl^2 N T}{3\pi^3}\int_{\sim 1}^{x/2}\frac{Q^4 dQ}{\left(  1+ \frac{\tl N Q^2}{3\pi^2}\ln \frac{\tL}{\sqrt{x^2 - 2 x Q}} \right)^2 + \left( \frac{\tl N Q^2}{6\pi} \right)^2} \approx \frac{\tl^2 N T}{3\pi^3}\int_{0}^{x/2} Q^4 dQ = \frac{\tl^2 N T x^5}{15 \cdot 2^5 \pi^3}.
\ee
We see that main contribution comes from $Q\le x/2.$

\subsubsection{2. Regime $ 1/\sqrt{\tl N \ln \tilde \Lambda} \ll x \ll \tL$}

Analogously to the previous case, main contribution in this regime comes from  the term with $\Omega_-$ in region 3. It is determined by momenta satisfying $Q\le x/2.$ After straightforward calculation, we find:

\be
\text{Im}\, \Sigma^A_{\ve}(x,0) = \frac{\tl^2 N T}{3\pi^3}\int_{\sim 1}^{x/2}\frac{Q^4 dQ}{\left(  1+ \frac{\tl N Q^2}{3\pi^2}\ln \frac{\tL}{\sqrt{x^2 - 2 x Q}} \right)^2 + \left( \frac{\tl N Q^2}{6\pi} \right)^2} \approx \frac{3\pi  T}{N}\int_{0}^{x/2} \frac{dQ}{\ln^2\frac{\tL}{\sqrt{x^2-2xQ}}} \approx \frac{3\pi T}{2N}\frac{x}{\ln^2\frac{\tL}x}.
\ee
This result, as before, is asymptotically correct provided $\ln (\tL/x) \gg1,$ since the subleading corrections have higher powers of the factor $1/\ln (\tL/x ).$

\subsection{E. Perturbative calculation \label{SMSec:Perturb}}


As we demonstrated above, the imaginary part of self-energy due to the Hubbard interaction can be described by the second-order perturbation theory in a wide range of energies/momenta, $T\exp(-b_2/\tl N)\ll \ve, \vf k \ll T\sqrt{1/\tl N \ln \tL}.$ In particular, it allows us to obtain the correct numerical prefactor in the regime $T\exp(-b_2/\tl N)\ll \ve, \vf k \ll T$.

Since the bare interaction is real, first non-vanishing contribution to $\text{Im} \, \Sigma$ is given by the diagram shown in Fig.~\ref{SMFig} (c), i.e., we keep only a single 'bubble' instead of the whole RPA series. Another (non-RPA) second-order diagram can be neglected in the large-$N$ limit which we assume in this paper. Imaginary part of the effective interaction then reads as

\be
\text{Im} V^R(\omega, \bq) = - V_0^2(\bq) N \text{Im}\Pi^R(\omega, \bq), \label{SMEq:Vperturb}
\ee
i.e., proportional to the imaginary part of the polarization operator. The latter can be expressed as

\begin{multline}
\text{Im} \, \Pi^R ( \omega, \bq) = \frac1{(2\pi)^{4}} \int d^3\bp \, d\ve \, \left( \tanh\frac{\ve}{2T} - \tanh\frac{\ve - \omega}{2T}   \right) \text{Tr} \,  G'' (\ve, \bp) G'' (\ve - \omega, \bp - \bq) = \\ =\frac{1}{2^5 \pi^2} \int d^3\bp \left\{ 2 \left[ 1 + \frac{\bp\cdot(\bp - \bq)}{p |\bp - \bq|}\right]\delta(\vf p - \omega - \vf |\bp - \bq| )\left[\tanh\frac{\vf p}{2T} - \tanh\frac{\vf |\bp - \bq|}{2T}  \right]+ \right. \\ \left. +  \left[ 1 - \frac{\bp\cdot(\bp - \bq)}{p |\bp - \bq|}\right]\left[ \delta(\vf p - \omega + \vf |\bp - \bq| ) - \delta(\vf p + \omega + \vf |\bp - \bq|) \right] \left[\tanh\frac{\vf p}{2T} + \tanh\frac{\vf |\bp - \bq|}{2T}  \right] \right\}, \label{SMEq:Piperturb}
\end{multline}
with $G'' = (G^R - G^A)/2i.$ Unlike our previous calculation, here we do not use elliptical coordinates and keep the expression for $\text{Im} \, \Pi^R$ in this form.

The imaginary part of the self-energy is given then by
\begin{multline}
\text{Im} \, \Sigma^R_{\ve}(\ve,\bk) = \frac14 \sum_{i = \pm} \int \frac{d^3q}{(2\pi)^3} \text{Im} V^R(\omega_i, \bq) \left(\coth\frac{\omega_i}{2T} + \tanh\frac{\ve - \omega_i}{2T}  \right) = \\ = -\frac{N}4 \sum_{i = \pm} \int \frac{d^3q}{(2\pi)^3}  V_0^2(\bq) \text{Im}\Pi^R(\omega, \bq) \left(\coth\frac{\omega_i}{2T} + \tanh\frac{\ve - \omega_i}{2T}  \right),
\end{multline}
with $\omega_{\pm} =\ve \pm \vf|\bk - \bq|.$ Using Eq.~(\ref{SMEq:Piperturb}), we find

\be
\text{Im} \, \Sigma^A_{\ve} (\bk, \ve) = \frac{\tl^2 N T}{2^5 \pi^{5}} f\left(\frac{\ve}{2T} , \frac{\vf k}{2T} \right),
\ee
where $f(\ve, k)$ expressed in dimensionless variables has form

\begin{multline}
f(\ve, k) = \int d^3\bp d^3\bq \left\{   \left[ \tanh q + \coth(\ve-q) \right] \left\{\left[ 1+  \frac{\bp \cdot(\bp - \bq - \bk)}{p|\bp - \bq - \bk|} \right]  \left[ \delta(p-\ve+q-|\bp - \bq - \bk|) - \right. \right. \right. \\ \left. \left. - \delta(-p - \ve + q + |\bp - \bq - \bk|) \right] \left[\tanh p - \tanh|\bp - \bq - \bk|  \right] + \left[ 1 -  \frac{\bp \cdot(\bp - \bq - \bk)}{p|\bp - \bq - \bk|} \right]  \left[ \delta(p-\ve+q + |\bp - \bq - \bk|) - \right. \right. \\ - \left. \left. \delta(p + \ve - q + |\bp - \bq - \bk|)\right] \left[ \tanh p + \tanh |\bp - \bq - \bk|  \right] \right\} + \\  \left. +  \left[ -\tanh q + \coth(\ve+q) \right] \left\{\left[ 1+  \frac{\bp \cdot(\bp - \bq - \bk)}{p|\bp - \bq - \bk|} \right]  \left[ \delta(p-\ve - q-|\bp - \bq - \bk|)  - \delta(-p - \ve - q + |\bp - \bq - \bk|) \right] \times \right. \right. \\  \left. \left.  \times\left[\tanh p - \tanh|\bp - \bq - \bk|  \right] + \left[ 1 -  \frac{\bp \cdot(\bp - \bq - \bk)}{p|\bp - \bq - \bk|} \right]  \left[ \delta(p-\ve - q + |\bp - \bq - \bk|) -  \delta(p + \ve + q + |\bp - \bq - \bk|)\right] \times \right. \right. \\ \left. \left. \times \left[ \tanh p + \tanh |\bp - \bq - \bk|  \right] \right\}\right\}.
\end{multline}
This expression can be significantly simplified in a number of important limits. In particular, for the cases when either energy or momentum is zero, we find

\begin{multline}
f(\ve, \bk = 0) = \int d\bp d\bq \left\{  2 \left[ 1 + \frac{\bp \cdot (\bp - \bq)}{p |\bp - \bq|} \right] \left[ \tanh p - \tanh|\bp - \bq|  \right]\left[ \tanh q + \coth(|\ve| - q)  \right] \delta ( p+q-|\bp - \bq| - |\ve|) +  \right.  \\ \left. +   \left[ 1 - \frac{\bp \cdot (\bp - \bq)}{p |\bp - \bq|} \right] \left[ \tanh p + \tanh|\bp - \bq|  \right]\left[\left[ \tanh q + \coth(|\ve| - q)  \right] \delta ( p+q + |\bp - \bq| - |\ve|) - \right. \right. \\ \left. \left. - \left[ \tanh q - \coth(|\ve| + q) \right] \delta ( p - q + |\bp - \bq| - |\ve|)\right] \right\},
\end{multline}

\begin{multline}
f(\ve = 0, \bk) = 4\int \frac{d\bp d\bq}{\sinh 2q}\left\{ 2\left[  1 + \frac{\bp \cdot (\bp - \bq - \bk)}{p |\bp - \bq - \bk|} \right] \left[ \tanh|\bp - \bq - \bk| - \tanh p \right] \delta(p + q - |\bp - \bq - \bk|)  + \right. \\ \left. +  \left[  1 - \frac{\bp \cdot (\bp - \bq - \bk)}{p |\bp - \bq - \bk|} \right] \left[ \tanh|\bp - \bq - \bk| + \tanh p \right] \delta(p - q + |\bp - \bq - \bk|)  \right\}.
\end{multline}

Finally, in the limit when both  energy and momentum are small compared to temperature, $|\ve|, \vf k \ll T$, we obtain

\begin{multline}
f(\ve, k) = 4\int \frac{d^3\bp d^3\bq}{\sinh 2q} \left\{ 2 \left[ \tanh(p + q) - \tanh p \right] \left[\delta(p + q - \ve - |\bp - \bq - \bk|) +  \delta(p + q + \ve - |\bp - \bq - \bk|)  \right]+  \right. \\ \left. + \left[ \tanh(q - p) + \tanh p \right]\left[ \delta(p - q + \ve + |\bp - \bq - \bk|)     +    \delta(p - q - \ve + |\bp - \bq - \bk|) \right] \right\}.
\end{multline}
It can be further shown that the result is universal if  $\ve,  k \to 0$ (does not depend on the ratio $\ve/ k$) and given by

\be
f(\ve \to 0, k \to 0) = 96 \pi^2 \int_0^{\infty} dp \int_0^{\infty} dq \frac{qp(p+q)}{\sinh 2q}\left[ \tanh(p+q) - \tanh p  \right] \approx 343.755.
\ee
This leads to

\be
\text{Im} \, \Sigma^A_{\ve} (\ve \to 0, \bk \to 0) = 0.0352\tl^2 N T.
\ee

\bibliographystyle{apsrev}

\end{document}